\documentclass[twocolumn, switch]{article} 
\usepackage{preprint}

\usepackage{amsmath, amsthm, amssymb, amsfonts}
\usepackage{apalike}

\usepackage[utf8]{inputenc}
\usepackage[T1]{fontenc}
\usepackage{xcolor}
\usepackage[hidelinks]{hyperref}
\usepackage{booktabs}
\usepackage{nicefrac}	
\usepackage{microtype}
\usepackage{lineno}
\usepackage{float}

\usepackage[hidelinks]{hyperref}
\usepackage[nolist,nohyperlinks]{acronym}
\usepackage{tabularx}
\usepackage{multirow}
\usepackage[table]{xcolor}
\usepackage{etoolbox}
\AtBeginEnvironment{quote}{\itshape}
\usepackage{url}

\usepackage{enumitem}
\setlist[itemize,enumerate]{leftmargin=1.8em, itemindent=0em}

\newcommand{\around}{\raise.17ex\hbox{$\scriptstyle\sim$}}

\begin{acronym}
    \acro{AI}{artificial intelligence}
    \acro{GPT}{generative pre-trained transformer}
    \acro{LLM}{large language model}
    \acro{NLP}{natural language processing}
    \acro{SNOW}{Steganographic Nature of Whitespace}
\end{acronym}

\usepackage{newfloat}
\DeclareFloatingEnvironment[name={Supplementary Figure}]{suppfigure}
\usepackage{sidecap}
\sidecaptionvpos{figure}{c}

\usepackage{titlesec}
\titlespacing\section{0pt}{12pt plus 3pt minus 3pt}{1pt plus 1pt minus 1pt}
\titlespacing\subsection{0pt}{10pt plus 3pt minus 3pt}{1pt plus 1pt minus 1pt}
\titlespacing\subsubsection{0pt}{8pt plus 3pt minus 3pt}{1pt plus 1pt minus 1pt}

\usepackage{tikz,xcolor,hyperref}

\definecolor{lime}{HTML}{A6CE39}
\DeclareRobustCommand{\orcidicon}{
	\begin{tikzpicture}
		\draw[lime, fill=lime] (0,0) 
		circle [radius=0.16] 
		node[white] {{\fontfamily{qag}\selectfont \tiny ID}};
		\draw[white, fill=white] (-0.0625,0.095) 
		circle [radius=0.007];
	\end{tikzpicture}
	\hspace{-2mm}
}
\foreach \x in {A, ..., Z}{\expandafter\xdef\csname orcid\x\endcsname{\noexpand\href{https://orcid.org/\csname orcidauthor\x\endcsname}
		{\noexpand\orcidicon}}
}

\title{Security and Detectability Analysis of Unicode Text Watermarking Methods Against Large Language Models}

\newcommand{\Footerhint}{\centering\footnotesize\color{gray!90}This is the author's version of a paper, accepted and presented at the \emph{ICISSP 2026} conference, published under the \href{https://creativecommons.org/licenses/by-nc-nd/4.0/}{CC BY-NC-ND 4.0} license. See \href{https://doi.org/10.5220/0014268700004061}{doi:10.5220/0014268700004061}}

\fancypagestyle{firstpage}{
	\fancyhf{}
	\fancyfoot[RE,LO]{\Footerhint}
}

\AtBeginDocument{\setlength{\footskip}{24pt}}

\usepackage{authblk}

\author[1]{Malte Hellmeier\orcidA{}}

\affil[1]{Fraunhofer ISST, Dortmund, Germany}

\begin{document}
	
\twocolumn[
\begin{@twocolumnfalse} 
    
    \maketitle
    \thispagestyle{firstpage}
    
    \begin{abstract}
        Securing digital text is becoming increasingly relevant due to the widespread use of large language models. Individuals' fear of losing control over data when it is being used to train such machine learning models or when distinguishing model-generated output from text written by humans. Digital watermarking provides additional protection by embedding an invisible watermark within the data that requires protection.
        However, little work has been taken to analyze and verify if existing digital text watermarking methods are secure and undetectable by large language models. In this paper, we investigate the security-related area of watermarking and machine learning models for text data. In a controlled testbed of three experiments, ten existing Unicode text watermarking methods were implemented and analyzed across six large language models: GPT-5, GPT-4o, Teuken 7B, Llama 3.3, Claude Sonnet 4, and Gemini 2.5 Pro. The findings of our experiments indicate that, especially the latest reasoning models, can detect a watermarked text. Nevertheless, all models fail to extract the watermark unless implementation details in the form of source code are provided. We discuss the implications for security researchers and practitioners and outline future research opportunities to address security concerns.
    \end{abstract}
    \keywords{Digital Text Watermarking \and Information Hiding \and Steganography \and Large Language Models \and Security Evaluation}
    \vspace{0.73cm}
    
\end{@twocolumnfalse} 
]

\section{Introduction}
\label{sec:introduction}
Securing and identifying textual data has become increasingly important in recent years, with an emerging focus on text used by \acp{LLM}. Big technology companies utilize massive public and private text data to train their models~\cite{Google.2025,OpenAI.2025}. In this regard, it has become increasingly complex to distinguish a text generated by an \ac{LLM} and a text written by a human~\cite{Kobis.2021}.

One possible solution to protect and identify text data lies in the use of watermarking techniques~\cite{Hellmeier.2023b}.
Existing techniques to protect digital text data date back to the 20th century, for example, techniques that append an encoded set of whitespaces at the end of a text to hide a secret message or watermark inside it~\cite{Kwan.2013}. Since then, various techniques have emerged, some focusing primarily on watermarks in \acp{LLM}~\cite{Liu.2025}. Due to the growing research interest, the security of these techniques is of utmost importance~\cite{Ahvanooey.2022}.

However, little attention has been paid to the detectability and security of watermarks when used within \acp{LLM}. Are existing state-of-the-art \acp{LLM} able to detect if a text contains a watermark? Are they able to extract the original watermark from a protected text? If such questions can be answered with \emph{yes}, the watermarking techniques fail their initial aim of safeguarding and securing text assets, as \acp{LLM} can identify, extract, or remove them.

To address this gap, the present study aims to evaluate ten existing text watermarking techniques across six \acp{LLM} to assess their detectability and security. Based on an extensive implementation testbed, we conducted three main experiments in a controlled setup to close this gap. Consequently, our work makes the following concrete contributions:

\begin{itemize}
    \item We evaluate the detectability of ten text watermarking algorithms in six different state-of-the-art \acp{LLM}.
    \item We evaluate the security and possibility of the same \acp{LLM} to extract and identify watermarks.
    \item We propose concrete suggestions for research and practice on how to improve the security of text watermarking.
\end{itemize}

\section{Background \& Related Work}
\label{sec:background}
The background information required to understand this work is introduced in the following section, along with related work on digital watermarking and \acp{LLM}.

\subsection{Digital Watermarking}
\label{subsec:watermarking}
The main idea of \emph{information hiding} lies in embedding data secretly inside other cover data, such as hiding information inside text, image, audio, or video~\cite{Ahvanooey.2018b}. Different subdomains have arisen under the umbrella of information hiding~\cite{Ahvanooey.2022}. One example is \emph{steganography}, which aims to hide data inside a cover in an invisible way, often used for secret communication~\cite{Petitcolas.1999}. In contrast, \emph{watermarking} embeds a secret message, known as a watermark, inside a cover and aims for robust embedding, copyright protection, and the prevention of illegal copying~\cite{Ahvanooey.2018b,Alkawaz.2016}. Therefore, the boundaries between steganography and watermarking are fluid. In this paper, we use the `watermark' term for unification purposes, since it fits the initially introduced problem domain.

To maintain clarity, this work analyzes existing \emph{format-based} methods that can hide a text watermark within another Unicode cover text by altering the format or layout of the text rather than its semantic content~\cite{Liu.2025,Zhang.2024}. Techniques such as generation-based or linguistic-based watermarking are excluded because they either generate entirely new content or heavily alter existing text~\cite{Ahvanooey.2022,Liu.2025,Zhang.2024}. Those techniques fail when applied to existing semantically sensitive texts, such as contracts, poems, or legal documents, where content changes can lead to issues~\cite{Alkawaz.2016}.

\subsubsection{Chosen Watermarking Algorithms}
\label{subsubsec:chosen-algorithms}
We utilized a set of ten related format-based text watermarking methods for our experiments, introduced in Table~\ref{tab:chosen-algorithms} and also used in related work~\cite{Hellmeier.2025c}. We focus on techniques that work with plain text, excluding format-based algorithms that change font attributes, such as type or color, since \acp{LLM} operate format-independently. Those techniques mainly make use of similar-looking characters, known as \emph{Unicode confusables} or homographs, which are defined as ``a letter or string that has enough of a visual similarity to a different letter or string that the two may be confused for one another''~\cite[p.~261]{Holgers.2006}. Other techniques use specific characters, such as whitespaces, control characters, or \emph{zero-width characters}, which are valid Unicode characters but invisible to humans because they have a displayed width of zero~\cite{TheUnicodeConsortium.2024}.

\begin{table*}[ht]
	\caption{Chosen Watermarking Algorithm Overview}
	\label{tab:chosen-algorithms}
	\centering
	\small
	\begin{tabularx}{\linewidth}{lp{35pt}p{288pt}p{100pt}}
		\hline
		Name & Type & Description & Source \\
		\hline
		AITSteg & Paper & Made for SMS by appending zero-width characters at the beginning of a cover. & \cite{Ahvanooey.2018} \\
		CovertSYS & Paper & Made for SMS by appending zero-width characters at the end of a cover, also using a one-time pad solution. & \cite{Ahvanooey.2022b} \\
		Innamark & Paper, Kotlin & One-to-one replacement of whitespaces with similar-looking whitespaces and options for compression, hashing, and error-correcting codes. & \cite{Hellmeier.2025c,Hellmeier.2025,FraunhoferISST.2025} \\
		LookALikes & Python & Replacement of 23 Unicode confusable letters using binary encoding. & \cite{Thompson.2021} \\
		Rizzo & Paper & Replacement of whitespaces and Unicode confusables, additionally protected with a secret password and hashing. & \cite{Rizzo.2016,Rizzo.2019} \\
		Sh.-Ur-Ra. & Paper & Replacement of whitespaces and Unicode confusables combined with an encryption technique.& \cite{ShazzadUrRahman.2021,ShazzadUrRahman.2023} \\
		Shiu & Paper & Made for social media messages by appending whitespaces at the end of lines and inserting additional line breaks. & \cite{Shiu.2018} \\
		\ac{SNOW} & C & Appending tabs and spaces at the end of a cover text. & \cite{Kwan.2013,Kwan.2016} \\
		StegCloak & Blog post, JavaScript & Zero-width characters embedded inside the cover with optional compression and encryption. & \cite{KuroLabs.2020,Mohanasundar.2020} \\
		UniSpaCh & Paper & Adding additional small whitespaces between words, sentences, and paragraphs. & \cite{Por.2012} \\
		\hline
	\end{tabularx}
\end{table*}

\subsection{Large Language Models}
\label{subsec:llms}
In our current \ac{AI} era, researchers and practitioners have begun building and training a range of models for various tasks. One major field of interest is \emph{\ac{NLP}} tasks involving human text and speech across various languages, where deep learning methodologies can be applied. The invention of the \emph{Transformer} architecture introduced dependencies between the input and output, enabling greater parallelization and faster model training~\cite{Vaswani.2017}. This transformer architecture has led to \emph{\acp{LLM}}, which are models trained on large text datasets to solve complex tasks by generating text~\cite{Zhao.2023b}. For these tasks, \emph{tokenization} is an essential step in \ac{NLP} because it converts text into tokenized units based on specific tokenization schemes~\cite{Webster.1992}.

Following the release of ChatGPT by OpenAI in 2022~\cite{OpenAI.2022}, an overwhelming interest in \acp{LLM} emerged in various domains among researchers, companies, and individuals. After the release of the first \ac{GPT} from OpenAI~\cite{Radford.2018} up to the latest \ac{GPT}-5.1 model released in 2025, various alternative \ac{LLM} models have arisen. In this work, we base our experiments on six widely used \acp{LLM}, which are introduced in the following.

\subsubsection{Chosen LLMs}
\label{subsubsec:chosen-llms}

We chose six different state-of-the-art \acp{LLM} for our experiments and analysis. While different \ac{LLM} operators have published minor variants of their models, such as mini, nano, or lite versions~\cite{Google.2025b,OpenAI.2025}, we selected the most powerful versions. All chosen \acp{LLM}, the versions used, and system cards are listed in Table~\ref{tab:chosen-llms} for reproducibility.

\begin{table}[ht]
    \caption{Chosen \acp{LLM} Overview}
    \label{tab:chosen-llms}
    \centering
    \begin{tabularx}{\linewidth}{lll}
        \hline
        Name & Version & System Card \\
        \hline
        GPT-5 & 2025-08-07 & \cite{OpenAI.2025} \\
        GPT-4o & 2024-11-20 & \cite{OpenAI.2024} \\
        Teuken 7B & instruct-v0.6 & \cite{openGPTX.2025} \\
        Llama 3.3 & 70B & \cite{Meta.2025} \\
        Claude Sonnet 4 & 20250514-v1 & \cite{Anthropic.2025} \\
        Gemini 2.5 & Pro & \cite{Google.2025} \\
        \hline
    \end{tabularx}
\end{table}

\section{Experimental Setup}
\label{sec:experimental-setup}
To validate the security and detectability, we employed all the watermarking algorithms introduced in Section~\ref{subsubsec:chosen-algorithms} and tested them on the current state-of-the-art \acp{LLM} described in Section~\ref{subsubsec:chosen-llms}.

Some of the algorithms have already been published and implemented, while the others are presented only in text form (cf. Table~\ref{tab:chosen-algorithms}). To build a valid testbed, we reimplemented all ten text watermarking algorithms in Java, based on their descriptions and existing source code. Some algorithms offer additional functionality, such as encryption techniques and additional passwords. Those functionalities encrypt the payload on top of the watermarking, while deactivating or activating them would not change the watermark-embedding process. Thus, we excluded these functionalities because we aim to test the detectability and security of the embedding layer in a comparable setup and do not want to distort the results with additional cryptographic features.

Regarding the \acp{LLM} used, we tested the models listed in Table~\ref{tab:chosen-llms}. We primarily used the default configuration for all models, setting the \texttt{reasoning\_effort}\footnote{Retrieved December 12, 2025, from \url{https://learn.microsoft.com/en-us/azure/databricks/machine-learning/model-serving/query-reason-models}} to \texttt{medium} and the \texttt{temperature}\footnote{Retrieved December 12, 2025, from  \url{https://learn.microsoft.com/en-us/azure/ai-foundry/openai/reference}} to \texttt{1.0} if the models support these parameters. Since all prompts act as user prompts, we do not configure a specific system prompt. To maintain compatibility between the \acp{LLM}, we used the same configuration and did not activate additional functionalities, such as the `Code execution' feature in Google AI Studio\footnote{Retrieved December 12, 2025, from \url{https://aistudio.google.com/}} for Gemini.

Our primary analysis is divided into three experimental parts, where each experiment uses a watermarked Lorem ipsum dummy text with an embedded `Secret Message` watermark inside it. For transparency and reproducibility, all prompts and the cover text used during the experimental runs are published online as supplementary material~\cite{Hellmeier.2025e}. The experiments form a three-tier attacker model, which is presented in the following.
\begin{itemize}
    \item \textbf{Tier A}: Uninformed attacker without prior watermarking knowledge (Experiment 1).
    \item \textbf{Tier B}: Partially informed attacker, knowing the used method (Experiment 2).
    \item \textbf{Tier C}: Well-informed attacker, having insider knowledge and code access (Experiment 3).
\end{itemize}

\paragraph{Experiment 1: Detectability.} The first experiment aims to test the detectability of a watermarked text. It assumes a Tier~A class of uninformed black-box attackers with no information about the watermarking method used. We asked every \ac{LLM} whether a given watermarked text contains a watermark. To prevent vague answers, we forced the \acp{LLM} to either respond with a \emph{Yes} if the text contains a watermark or a \emph{No} if it does not. Our evaluation clusters the results by \ac{LLM} and by algorithm accuracy, with \emph{Yes} indicating correct answers for watermarked inputs and \emph{No} for the original input text without a watermark.

\paragraph{Experiment 2: Security with Name.} The second experiment aims to test the security of a watermarked text by assessing whether \acp{LLM} can detect and extract the watermark. It assumes a Tier~B class of attackers who are aware of the watermarking method used. Therefore, we asked each \ac{LLM} to extract the watermark by providing the watermarked text and the name of the algorithm or the name of the publication, along with its authors, if no specific algorithm name existed. Similar to the previous experiment, we prevent vague answers by forcing the \acp{LLM} to either respond with the extracted watermark or with the word \emph{Unsure} if an extraction is not possible. Our evaluation classifies successful outcomes as those in which the full `Secret Message` watermark is extracted, while ignoring additional explanations. Extractions like `secret` only are classified as partial matches, and any other results are considered incorrect and as unsuccessful extractions.

\paragraph{Experiment 3: Security with Code.} The last experiment extends the previous security experiment by additionally providing the source code of the used watermarking algorithm in the user prompt as model input. It assumes a Tier~C class of attackers, possessing in-depth knowledge of the watermarking algorithm used. Each chosen watermarking algorithm is implemented in our testbed as a single Java class that adheres to a consistent structure. Consequently, every class has an \texttt{addWatermark(...)} method to embed a watermark inside a cover and a \texttt{getWatermark(...)} method to extract a watermark from a given text. Our evaluation uses the same metrics for successful and unsuccessful extractions as in the previous Experiment 2.

\section{Results}
\label{sec:results}
The results of each of the three experiments, including identified errors, are presented below and interpreted and explained in more detail in the upcoming discussion of Section~\ref{sec:discussion}. The complete result set of all \ac{LLM} responses is published online as supplementary material for transparency and further analysis~\cite{Hellmeier.2025e}.

\paragraph{General Errors.} Even if we explicitly tell the \acp{LLM} to only answer with specific words like \emph{Yes},  \emph{No}, \emph{Unsure}, or the extracted watermark, they sometimes also answer differently. We observe two classes of errors. First, different non-compliant answers in another format, like responses with longer text or explanations. Second, the \acp{LLM} believe they have identified the correct watermark, but respond with a wrong String that does not match our `Secret Message' watermark, classified as unsuccessful extraction.

\begin{figure}[ht]
    \centering
    \includegraphics[width=0.9\columnwidth]{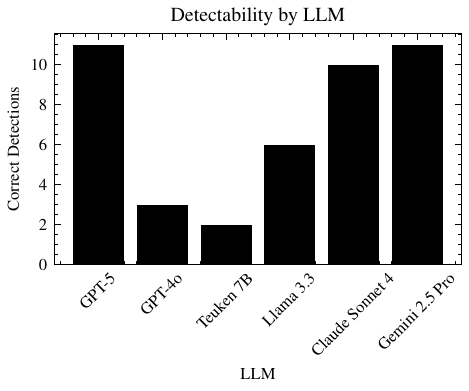}    
    \caption{Detectability Analysis by \ac{LLM} (Experiment 1)}
    \label{fig:detectability-llm-results}
\end{figure}

\paragraph{Experiment 1: Detectability.} The first experiment tested whether the \acp{LLM} could detect that a given text contains a watermark. Figure~\ref{fig:detectability-llm-results} shows the results of all ten tested watermarking algorithms, along with the original unwatermarked text, grouped by \ac{LLM}. Here, \ac{GPT}-5 and Gemini 2.5 Pro correctly detected all cases, while Claude Sonnet 4 was not able to detect the Innamark watermark, but correctly detected all other cases. The results were significantly worse for the other three \acp{LLM}, whereas Teuken 7B could only detect the original unwatermarked text and the text watermarked by the StegCloak algorithm.

\begin{figure}[ht]
    \centering
    \includegraphics[width=0.9\columnwidth]{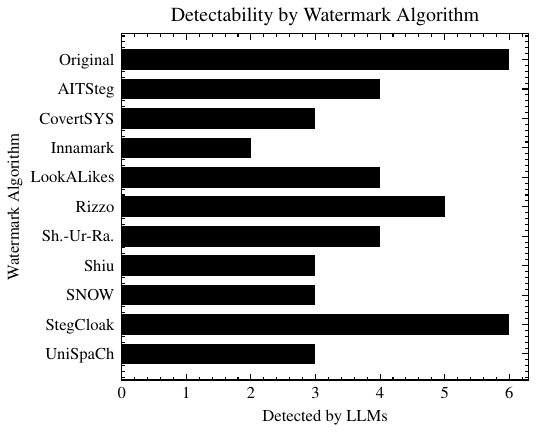}
    \caption{Detectability Analysis by Watermarking Algorithms (Experiment 1)}
    \label{fig:detectability-watermark-results}
\end{figure}

Figure~\ref{fig:detectability-watermark-results} shows the results of Experiment 1 grouped by the watermarking algorithms. All six tested \acp{LLM} were able to detect that the text watermarked by the StegCloak algorithm contains a watermark and that the original text does not contain any watermark. The Innamark algorithm shows the best imperceptibility because it is only detected by GPT-5 and Gemini 2.5 Pro, but goes unnoticed for the other four \acp{LLM}.

\paragraph{Experiment 2: Security with Name.} The second experiment asks the \acp{LLM} to extract the watermark by explicitly mentioning the watermarking algorithm used. The results are shown in Table~\ref{tab:extracted-watermarks-with-name}, color-coded with red for unsuccessful, yellow for partly successful, and green for successful results. As shown in the table, none of the \acp{LLM} successfully extracted the full `Secret Message' watermark in any case. Only Gemini 2.5 Pro was able to extract `this is a secret message' from the watermarked text by StegCloak and `secret' from the UniSpaCh text. Furthermore, it is noticeable that Teuken 7B answered only four times in the given structure with the word \emph{Unsure}. In the other six cases, the instructions were not followed; instead, an apology was issued, and a more detailed explanation was given. Furthermore, Llama 3.3 incorrectly assumed that the names of algorithms or authors were used as watermarks in three cases. The majority of results are mostly only \emph{Unsure}, with a few exceptions such as `A1' in a Gemini 2.5 Pro case, `Hello' in a Claude Sonnet 4 case, `Hello Im Your Father' in a Llama 3.3 case, and `ZEZRGEPi' in a GPT-5 case, as can be inferred from Table~\ref{tab:extracted-watermarks-with-name}.

\begin{table*}[ht]
    \caption{Extracted Watermarks by the \acp{LLM} with Name (Experiment 2)}
    \label{tab:extracted-watermarks-with-name}
    \centering
    \begin{tabularx}{\textwidth}{l|lllllp{99pt}}
        \hline
        & GPT-5 & GPT-4o & Teuken 7B & Llama 3.3 & Claude Sonnet 4 & Gemini 2.5 Pro\\
        \hline
        AITSteg & \cellcolor{red!20}ZEZRGEPi & \cellcolor{red!20}Unsure & \cellcolor{red!20}Unsure & \cellcolor{red!20}Unsure & \cellcolor{red!20}Unsure & \cellcolor{red!20}Unsure\\
        CovertSYS & \cellcolor{red!20}Unsure & \cellcolor{red!20}Unsure & \cellcolor{red!20}I'm sorry, ... & \cellcolor{red!20}Unsure & \cellcolor{red!20}Unsure & \cellcolor{red!20}Unsure \\
        Innamark & \cellcolor{red!20}Unsure & \cellcolor{red!20}Unsure & \cellcolor{red!20}I'm sorry, ... & \cellcolor{red!20}Innamark & \cellcolor{red!20}Unsure & \cellcolor{red!20}Unsure \\
        LookALikes & \cellcolor{red!20}Unsure & \cellcolor{red!20}Unsure & \cellcolor{red!20}I'm sorry, ... & \cellcolor{red!20}Hello Im Your Father & \cellcolor{red!20}Unsure & \cellcolor{red!20}A1 \\
        Rizzo & \cellcolor{red!20}Unsure & \cellcolor{red!20}Unsure & \cellcolor{red!20}Unsure & \cellcolor{red!20}Rizzo, Stefano ... & \cellcolor{red!20}Unsure & \cellcolor{red!20}Unsure \\
        Sh.-Ur-Ra. & \cellcolor{red!20}Unsure & \cellcolor{red!20}Unsure & \cellcolor{red!20}I'm sorry, ... & \cellcolor{red!20}Shazzad-Ur-Ra... & \cellcolor{red!20}Unsure & \cellcolor{red!20}Unsure \\
        Shiu & \cellcolor{red!20}Unsure & \cellcolor{red!20}Unsure & \cellcolor{red!20}I'm sorry, ... & \cellcolor{red!20}Unsure & \cellcolor{red!20}Hello & \cellcolor{red!20}Unsure \\
        \ac{SNOW} & \cellcolor{red!20}Unsure & \cellcolor{red!20}Unsure & \cellcolor{red!20}Unsure & \cellcolor{red!20}Unsure & \cellcolor{red!20}Unsure & \cellcolor{red!20}Unsure \\
        StegCloak & \cellcolor{red!20}Unsure & \cellcolor{red!20}Unsure & \cellcolor{red!20}Unsure & \cellcolor{red!20}Unsure & \cellcolor{red!20}Unsure & \cellcolor{yellow!20}this is a secret message \\
        UniSpaCh & \cellcolor{red!20}Unsure & \cellcolor{red!20}Unsure & \cellcolor{red!20}I'm sorry, ... & \cellcolor{red!20}Unsure & \cellcolor{red!20}Unsure & \cellcolor{yellow!20}secret\\
        \hline
    \end{tabularx}
\end{table*}

\paragraph{Experiment 3: Security with Code.} The last experiment extends Experiment 2 by also passing our Java source code in the prompt to test security against extraction. Here, we need to exclude the Teuken~7B~v0.6 model from Experiment 3 because it has a maximum input size of 4,096 tokens and cannot process the long input prompt.

Table~\ref{tab:extracted-watermarks-with-code} shows the outputs of all execution runs in this experiment, using the same red-yellow-green color-coding. In Experiment 2, none of the \acp{LLM} fully extracted the watermark. However, the source code helps the models understand the structure and process, enabling them to extract the watermark partially. Since GPT-4o, Llama 3.3, and Claude Sonnet 4 were unable to successfully extract a watermark in any case, GPT-5 and Gemini 2.5 Pro successfully extracted the correct `Secret Message' watermark in some cases. Moreover, Claude Sonnet 4, Gemini 2.5 Pro, and GPT-5 have cases where they can extract parts of the watermark, such as the word `Secret' alone. It is further noticeable that Claude Sonnet 4 incorrectly assumes, in seven cases, that the watermark is a version of `Hello' or `Hello World'.

\begin{table*}[ht]
    \caption{Extracted Watermarks by the \acp{LLM} with Code (Experiment 3)}
    \label{tab:extracted-watermarks-with-code}
    \centering
    \begin{tabularx}{\textwidth}{l|ll>{\centering\arraybackslash}p{45pt}p{80pt}ll}
        \hline
        & GPT-5 & GPT-4o & Teuken 7B & Llama 3.3 & Claude Sonnet 4 & Gemini 2.5 Pro\\
        \hline
        AITSteg & \cellcolor{red!20}Unsure & \cellcolor{red!20}Unsure & \cellcolor{red!20}  & \cellcolor{red!20}Unsure & \cellcolor{red!20}Hello & \cellcolor{green!20}Secret Message \\
        CovertSYS & \cellcolor{red!20}Unsure & \cellcolor{red!20}Unsure & \cellcolor{red!20} & \cellcolor{red!20}Unsure & \cellcolor{red!20}Hello World & \cellcolor{yellow!20}Secret!! \\
        Innamark & \cellcolor{green!20}Secret Message & \cellcolor{red!20}Unsure & \cellcolor{red!20} & \cellcolor{red!20}Innamark & \cellcolor{red!20}Unsure & \cellcolor{green!20}Secret Message \\
        LookALikes & \cellcolor{yellow!20}Secret & \cellcolor{red!20}Unsure & \cellcolor{red!20} & \cellcolor{red!20}Hello & \cellcolor{red!20}Hello World! & \cellcolor{green!20}Secret Message \\
        Rizzo & \cellcolor{red!20}Unsure & \cellcolor{red!20}Unsure & \cellcolor{red!20} & \cellcolor{red!20}Watermarking extraction failed. & \cellcolor{red!20}Hello & \cellcolor{yellow!20}secret message \\
        Sh.-Ur-Ra. & \cellcolor{green!20}Secret Message & \cellcolor{red!20}Unsure & \cellcolor{red!20} & \cellcolor{red!20}Unsure & \cellcolor{red!20}Hello World & \cellcolor{green!20}Secret Message \\
        Shiu & \cellcolor{green!20}Secret Message & \cellcolor{red!20}HelloWorld! & \cellcolor{red!20} & \cellcolor{red!20}Unsure & \cellcolor{red!20}Hello World! & \cellcolor{red!20}SlVfHWa:I1ngdee \\
        \ac{SNOW} & \cellcolor{red!20}Unsure & \cellcolor{red!20}Unsure & \cellcolor{red!20} & \cellcolor{red!20}SNOW & \cellcolor{yellow!20}Secret & \cellcolor{red!20}Unsure \\
        StegCloak & \cellcolor{green!20}Secret Message & \cellcolor{red!20}Unsure & \cellcolor{red!20} & \cellcolor{red!20}Unsure & \cellcolor{yellow!20}Secret & \cellcolor{red!20}Unsure \\
        UniSpaCh & \cellcolor{green!20}Secret Message & \cellcolor{red!20}Unsure & \multirow{-2}{*}[10em]{\cellcolor{red!20}\rotatebox[origin=c]{90}{Input token length too short. }} & \cellcolor{red!20}Hello World & \cellcolor{red!20}Hello World! & \cellcolor{red!20}Unsure \\
        \hline
    \end{tabularx}
\end{table*}

\section{Discussion}
\label{sec:discussion}
The conducted experiments in this work analyzed the detectability and security of ten existing Unicode-based text watermarking systems across six of the latest \acp{LLM}. The now-daily use of various \acp{LLM} for different tasks highlights the need to scientifically validate whether existing watermarking mechanisms can still ensure protection and security, or whether they will become obsolete as \acp{LLM} learn to recognize and extract them. Our conducted experiments help to answer the open points. In the following, we interpret our experimental results by deriving implications for both theory and practice and discussing the limitations and future research opportunities.

\subsection{Theoretical and Practical Implications}
\label{subsec:implications}
Experiment 1 demonstrates that the latest models with reasoning functionality, such as GPT-5 and Gemini 2.5 Pro, can consistently detect and distinguish watermarked text from the original text in all cases. It is reasonable to assume that such Unicode-based text watermarking techniques may not be suitable for practical use cases where invisibility and unrecognizability are of great importance. A malicious attacker could process all text generated by an \ac{LLM} to determine whether it contains a watermark.

However, extracting the original watermark was impossible in all cases, as shown in Experiment 2. Even if the malicious attacker knows the watermarking method, all tested \acp{LLM} fail to extract the original watermark in all cases. Consequently, practitioners can still use those watermarking techniques without fear that the original watermark will be quickly extracted. We hypothesize that \acp{LLM} will be able to successfully extract watermarks in the near future as research and development advance. Our Experiment 3 has already shown that, with the implemented source code, some \acp{LLM} can understand and extract watermarks. Nevertheless, practitioners can apply upstream encryption techniques, such as simple passwords or asymmetric encryption with public and private keys, to mitigate the substantial risks of unauthorized extractions.

It is further noticeable that some of the tested \acp{LLM} tend to respond in a specific way. For example, the Teuken 7B model has issues following the initial instructions to only respond with `Unsure' or the extracted watermark, as shown in Experiment 2 (cf. Table~\ref{tab:extracted-watermarks-with-name}). Further, the Claude Sonnet 4 often responded with versions of `Hello World' in Experiment 3 (cf. Table~\ref{tab:extracted-watermarks-with-code}). It is conceivable that this is related to the models' training and structure. Since Claude Sonnet 4 specializes in coding tasks~\cite{Anthropic.2025}, it might explain the trend to `Hello World' answers because it is an often-used example string in software development. This leads to specific implications for researchers to review and refine existing models to prevent such a flawed bias.

\subsection{Limitations and Future Research}
\label{subsec:limitations-future-research}
Nevertheless, our conducted research is not free of limitations, as discussed in this section, along with concrete suggestions for future research.

First, our experimental setup analyzes \ac{LLM} outputs using a well-defined set of user prompts in a controlled environment with consistent settings. Nevertheless, \acp{LLM} are non-deterministic, which might lead to a different response if the same prompt is asked again in the same setup. Therefore, the results should not be considered individually but rather indicate a first snapshot of detectability and security. Future research is needed to run multiple trials with the same setting and different cover texts and watermarks. This would allow us to derive more stable statistical results, including quantitative indicators, by aggregating the results in a confusion matrix.

Second, we configured all parameters across the \acp{LLM} similarly to ensure comparable results. Nevertheless, parameter tuning can potentially improve detectability and accuracy of watermark extraction, for example, by reducing the temperature, given the deterministic nature of our tasks. Future work should analyze the influence of the different parameters to identify a configuration that achieves the best possible results.

Third, we implemented the watermarking algorithms in a Java testbed and built our analysis upon it. We validated the correctness of our implementations with different manual tests, but it would benefit from a logical analysis.

Fourth, we evaluated ten text watermarking algorithms across six \acp{LLM} using three attack experiments. Future research should extend this setup, including (1) other text watermarking methods besides format-based techniques, especially those used by \acp{LLM}~\cite{Liu.2025}; (2) other \acp{LLM}, such as the Claude Opus 4 model, which offers additional capabilities~\cite{Anthropic.2025}; or (3) additional attacks, such as paraphrasing, modifications, or tasking the \acp{LLM} to remove the watermark.

\section{Conclusions}
\label{sec:conclusion}
We have analyzed the security and detectability of current watermarking mechanisms for digital text. We have implemented ten text watermarking algorithms in a Java testbed and analyzed them across three controlled experiments using six state-of-the-art \acp{LLM}. While the detectability and security of text watermarks are underresearched, our results fill the gap by evaluating how current \acp{LLM} can detect and extract watermarks from protected text data.

We have found that \acp{LLM} are already able to detect and distinguish between watermarked and unwatermarked texts. Nevertheless, they still struggle to extract the original watermark without additional information about the initial watermark-embedding process. While our results are limited to a single input text, a single watermark payload, and a single execution run with default settings, future research is needed to expand our analysis to other watermarking mechanisms and \acp{LLM} with optimized configurations. Furthermore, both watermarking researchers and \ac{LLM} operators need to act: the former should develop mechanisms that are undetectable and secure when a watermarked text is used in \acp{LLM}, and the latter should improve their models to better detect and extract watermarks.

\footnotesize
\section*{Acknowledgements}
This research was supported by the Center of Excellence Logistics and IT, funded by the Fraunhofer-Gesellschaft.
We also thank Ernst-Christoph Schrewe for his efforts in supporting the development process of the watermarking testbed. 
During the preparation of this work, the author utilized AI-assisted tools such as Grammarly, DeepL, and OpenAI GPT models to enhance English language accuracy, including spelling, grammar, and punctuation. The content generated or corrected by these tools was thoroughly reviewed, revised, and edited by the author to ensure accuracy and originality. The author accepts full responsibility for the originality of the final content of this publication.

\bibliographystyle{apalike}
{\small
\bibliography{literature-arXiv}}
	
\end{document}